# Sensitivity analysis of GSI based mechanical characterization of rock mass

P. Ván[1] and B. Vásárhelyi[2]

[1]Dept. of Theoretical Physics, Wigner RCP, of the HAS and
Dept. of Energy Engineering, BME and
Montavid Thermodynamic Research Group, Budapest, Hungary
[2]Dept. of Structural Engng., Pollack Mihály Faculty of Engng., Univ. of Pécs, Pécs, Hungary

**Abstract**

Recently, the rock mechanical and rock engineering designs and calculations are frequently based on Geological Strength Index (GSI) method, because it is the only system that provides a complete set of mechanical properties for design purpose. Both the failure criteria and the deformation moduli of the rock mass can be calculated with GSI based equations, which consists of the disturbance factor, as well. The aim of this paper is the sensitivity analysis of GSI and disturbance factor dependent equations that characterize the mechanical properties of rock masses. The survey of the GSI system is not our purpose. The results show that the rock mass strength calculated by the Hoek-Brown failure criteria and both the Hoek-Diederichs and modified Hoek-Diederichs deformation moduli are highly sensitive to changes of both the *GSI* and the *D* factor, hence their exact determination is important for the rock engineering design.

## 1. Introduction

The sensitivity of different empirical formulas to parameter uncertainty is an important factor for a rock engineering designer. The purpose of this paper is to determine the sensitivity of the different mechanical equations based on the Geological Strength Index (GSI) and disturbance factor (D). Recently, Bieniawski (2011) demonstrates the high sensitivity of the Hoek-Brown failure criteria according to the results of Malkowski (2010): he shows that a change of 5 in the GSI value, from 35 to 40, leads to dramatic increases in the values of the following parameters: $\sigma_{cm}$ by 37%, change in parameter $m_b$ by 20% and in the modulus of deformation $E_M$ by 33%, while that of parameter *s* by 85%.

In order to establish good empirical formulas one should have some idea about the effect of variations in the input parameters for judging the acceptability of the design. Accordingly, we analyze the generalized Hoek-Brown formula and the Hoek-Diederichs and modified Hoek-Diederichs formulas for deformation modulus from this point of view, and give some practical tools for rapid sensitivity analyses. The first steps of this analysis were carried out by Ván and Vásárhelyi (2007).

## 2. Geological Strength Index (GSI) and the disturbance factor (D)

The Geological Strength Index (GSI), as a system of rock mass characterization, was introduced by Hoek (1994) and Hoek et al, (1995) and recently it is widely used in rock engineering designs. The goal of this engineering geological system was to present input data, particularly those related to rock mass properties required as inputs into numerical analysis or closed form solutions for designing tunnels, slopes or foundations in or on rocks. It provides a field method, so the geological character of rock material, together with the visual assessment of the mass it forms is used as a direct input to the selection of parameters relevant for the prediction of different mechanical parameter of the rock mass. This approach enables a rock mass to be considered as a mechanical continuum. Marinos et al (2005) review the application and the limitation of the Geological Strength Index, showing the deterimation methods. However, it is well known that the determination of this parameter is not easy and is not exact; it is encumbered by several uncertainties. On Figure 1 the general chart for GSI determination is presented according to Hoek and Marinos (2000). According to its definition "From the lithology, structure and surface conditions of the discontinuities, estimate the average value of GSI. Do not try to be too precise. Quoting a range from 33 to 37 is more realistic than starting that GSI = 35" (Hoek et al., 1992). Therefore, in relative terms the GSI here is 35±10% and because the exactness is given in absolute terms, for lower values the relative error increases. This is what is suggested using GSI in case of very weak and sheared rock masses, i.e. flysch and schist, where GSI < 30 (Marinos and Hoek, 2001 and Hoek et al, 1998, respectively). E.g. if the GSI = 10, (2 < GSI < 12) the sensitivity of this value reaches the 20 %! Also with the more exact methods for the calculation of the GSI value (see Sonmez and Urusay, 1999; Cai et al. 2004; and Russo, 2009) there are several possibilities of errors.

The influence of blast damage on the near surface rock mass properties have been taken into account in the 2002 version of the Hoek-Brown criterion (Hoek et al., 2002). *D* is a factor which depends upon the degree of disturbance due to blast damage and stress relaxation. It varies from 0 for undisturbed in situ rock masses to 1 for very disturbed rock masses. Guidelines for the selection of *D* are presented in Table 1. One can see, that the exact determination of the disturbance factor *D* is difficult – up to now it is not standardized. There are no guidelines except this one from the first version of Hoek et al (2002).

According to these guidelines 10-20% errors are tolerable. E.g. the good blasting $D = 0.7$, poor blasting $D = 1$ difference makes possible a $D = 0.8\pm0.1$ value with a 12.5% uncertainty in $D$.

## 3. Mechanical equations based on GSI and D values

Based on the GSI and disturbance factor ($D$) there are several formulas to calculate the failure and deformation moduli of the rock mass. These equations are presented below, which are based on empirical results, not any theoretical calculations:

*3.1 Hoek-Brown failure criterion*

The Hoek-Brown equation is one of the most popular failure criteria for determining the failure envelope of the rock mass. For jointed rock masses it is given by the following generalized formula (Hoek et al., 2002 and Eberhard, 2012):

$$\sigma_1' = \sigma_3' + \sigma_{ci}\left(m_b \frac{\sigma_3'}{\sigma_{ci}} + s\right)^a, \quad (1)$$

where

- $\sigma_1'$ and $\sigma_3'$ are the maximum and minimum effective principal stresses at failure;
- $\sigma_{ci}$ is the uniaxial compressive strength of the intact rock sections;
- $m_b$ is the value of the Hoek-Brown constant for the rock mass, depending on the Hoek-Brown constant of the intact rock ($m_i$), the Geological Strength Index (*GSI*) and the blast disturbance (*D*):

$$m_b = m_i \exp\left(\frac{GSI - 100}{28 - 14D}\right) \quad (2)$$

- $s$ and $a$ are parameters that also depend on the rock mass characteristics:

$$s = \exp\left(\frac{GSI - 100}{9 - 3D}\right) \quad (3)$$

and

$$a = \frac{1}{2} + \frac{1}{6}\left(e^{-GSI/15} - e^{-20/3}\right) \quad (4)$$

According to the Hoek-Brown equation (1) the ratio of the uniaxial compressive strength of the rock mass ($\sigma_{cm}$) and to that of the intact rock ($\sigma_{ci}$) can be determining:

$$\sigma_{cm}/\sigma_{ci} = s^a \tag{5}$$

Where $s$ and $a$ can be calculated by Eq. 3 and 4, respectively.

3.2 *Deformation modulus of rock mass*

The formula, introduced by Hoek and Diederichs (2006), calculates the deformation modulus from the *GSI* value and *D* factor as:

$$E_{rm}(MPa) = 100 \frac{1 - D/2}{1 + e^{(75 + 25D - GSI)/11}} \tag{6}$$

or if the deformation modulus of the intact rock ($E_i$) is known, equation (1) can be modified to:

$$E_{rm}(MPa) = E_i \left( 0.02 + \frac{1 - D/2}{1 + e^{(60 + 15D - GSI)/11}} \right) \tag{7}$$

Using the two formulas the estimated deformation moduli are not the same, they depend on the deformation modulus of the intact rock.

The uncertainty in the determination of *GSI* and *D* values has an additional interpretational subtlety in the light of the different empirical formulas. For example the *GSI* dependence of Eqs. (6) and (7) is qualitatively similar, as one can see on Figures 2 and 3. However, the corresponding values of deformation modulus can be very different. The ratio of the two values multiplied by the intact rock deformation modulus is plotted as the function of *GSI* on Figure 4 with disturbance factors ($D = 0$, 0.5 and 1), respectively. If the two formulas with identical *GSI* and *D* values were related to the same deformation modulus, then the plotted ratio should have been constant. One can see, that it increases when *GSI* runs form 0 to 100 at about 20 times in case of D = 0 and at about 200 times if $D = 1$. Therefore the *GSI* and also the *D* values of the same rock mass have to be interpreted and calculated differently depending on the applied formula to obtain the same deformation modulus.

**4. Sensitivity analysis**

The sensitivity of a function *f* regarding the uncertainties of the variables can be characterized by the formula commonly known as propagation of uncertainty or propagation of error (Bronstein & Semendjajew, 2004). Let us suppose that *f* is a real function which depends on *n* random variables $x_1$, $x_2$, … $x_n$. From their uncertainties $\Delta x_1, \Delta x_2, \ldots \Delta x_n$ we can calculate the uncertainty $\Delta f$ of *f* :

$$\Delta f = \left( \sum_{i=1}^{n} \left( \left. \frac{\partial f}{\partial x_i} \right|_{x_1,\ldots,x_{i-1},x_{i+1},\ldots,x_n} \Delta x_i \right)^2 \right)^{\frac{1}{2}} \tag{8}$$

Here it is assumed that the variables are uncorrelated and the underlying probability distribution of the errors is Gaussian.

Therefore if the variables $x_i$ are measured with an experimental error, $x_i \pm \Delta x_i$, we can estimate the uncertainty of their arbitrary function with the above formula. This formula is robust; the Gaussian distribution is a reasonable assumption in most cases. If the variables are correlated we should apply a modified equation for sensitivity estimates.

In this paper the relative sensitivity of the Hoek-Brown parameters, the rock mass strength and the deformation moduli of the rock mass were calculated in case of 5% and 10% relative uncertainties, that is when both $\Delta D/D$ and $\Delta GSI/GSI$ is 0.05 and when both $\Delta D/D$ and $\Delta GSI/GSI$ are 0.1, for *D* = 0; 0.5 and 1.0.

## 5. Results of the sensitivity analyses

- *Analysis of the sensitivity of the $m_b$ value*

The dependence of *GSI* on the ratio of the $m_b/m_i$ is plotted in Figure 5 in the case of 0; 0.5 and 1.0 values of disturbance factor *D*. The 5 % and 10 % *GSI* deviations were calculated and presented in Figures 6 and 7, respectively. We can see that the relative sensitivity of $m_b$ is at least double the uncertainties of the *GSI* and *D* values, and may be 7 times higher in case of large disturbance parameters and low and high GSI values.

- *Analysis of the sensitivity of s*

The dependence of *GSI* on the ratio of the *s* parameter is plotted in Figure 8, in case of 0; 0.5 and 1.0 values of disturbance factor *D*. Figures 9 and 10 show that the relative sensitivity of the s parameter is at least the triple of the uncertainties of the variables, and may even be 15 times higher (!) in the case of large disturbance parameters and high *GSI* values.

- *Analysis of the sensitivity of the a parameter*

The *a* parameter is independent of the disturbance factor and not sensitive to the uncertainties in GSI (Eq. 4, Figure 11). The maximum relative sensitivity of s is about equal to the uncertainty of the variables at GSI value 20. The relative sensitivity of *a* in the case of 5 % and 10 % measurement errors are plotted in Figure 12 and 13, respectively.

Finally, in Figure 14 the Hoek-Brown failure envelope is presented in 3D visualization (Eq. 1) and the sensitivity of this criteria is plotted in Figure 15 in case of 10 % errors (i.e.: $GSI\pm0.1GSI$ and $D\pm0.1D$).

- *Analysis of the sensitivity of the strength of the rock mass*

The dependence of *GSI* on the rock mass strength $\sigma_1$ (see Eq. (1)) in the case of various disturbance factors *D* is presented in Figure 16. According to Figures 17-18 at low *GSI* values the uncertainty in the disturbance parameter *D* determines the sensitivity of the rock mass strength, at high *GSI* values the uncertainty in *GSI* dominates and the disturbance parameters have less influence. Figures 19-20 show that the relative sensitivity of the rock mass strength $\sigma_1$ is at least double of the uncertainties in the *GSI* and disturbance parameter, and may be 8 times higher in case of large disturbance parameter and high *GSI* values.

- **Sensitivity analysis of the Hoek-Diederichs formulas**

The relative sensitivity for the simple Hoek-Diederichs equation (6) is plotted as a function of GSI in the case of 5 % relative uncertainty both in *GSI* and *D* in Figure 21 for disturbance values $D = 0$, 0.5 and 1. One can see that the sensitivity in the rock mass deformation modulus is between 15-35% and strongly depends on the GSI value. There is a peak in the sensitivity between GSI values of 60 and 80. Figure 22 shows the corresponding relative sensitivity according to the modified Hoek-Diederichs formula, Eq. 7. Here we assumed that the deformation modulus of the intact rock, $E_i$, is exact. The deformation modulus may change from 0.5 to 22% depending on the GSI value. The sensitivity of the modified Hoek-Diederich formula is independent of the intact rock deformation modulus. The peaked property is even more apparent in this case, with the greatest sensitivity occurring for *GSI* values between 40 and 60. Figures 23 and 24 show the similar curves with 10% relative uncertainty of the GSI and D values.

## 4. Discussion

The sophisticated empirical Hoek-Brown formula is sensitive to the uncertainties of the *GSI* and disturbance parameter (*D*) values. Its relative sensitivity may reach a value 8 times higher than the relative uncertainties of the *GSI* and *D* factors in the case of high disturbance and *GSI* values, if these relative uncertainties are uniform. With more exact *GSI* determination at high *GSI* values and disturbance factor (*D*) determination at low *GSI* values, the relative sensitivity of the Hoek-Brown formula can be considerably reduced.

The Hoek-Diederichs equations can enlarge the uncertainties of *GSI* and *D* up to seven times, the modified Hoek-Diederichs formula up to four times, depending on the *GSI* and *D* parameters. Here one can reduce the sensitivity of the equations by more exact determination in case of high disturbance factors and *GSI* in between 60 and 80 in case of Eq. 6. The modified formula Eq. 7 is most sensitive for GSI values between 20-60 for small *D* and *GSI* values between 50-90 for large D.

According to our analysis the Hoek-Brown failure criteria and the Hoek-Diederichs formulas can be highly sensitive to the uncertainties in the GSI and disturbance parameters. This sensitivity is due to the complex structure of the functions, criteria containing a lower number of parameters may be less sensitive. In any case the rock engineering design should consider the uncertainties of the design parameters and calculate them routinely. According to these results using the GSI system without any control is not recommended. Recently, similar results were found by Anagnostou and Pimentel (2012).

*Acknowledgements*
P. Ván acknowledges the financial support of the OTKA K81161, K104260 and TT 10-1-2011-0061/ZA-15-2009 grants for this research.

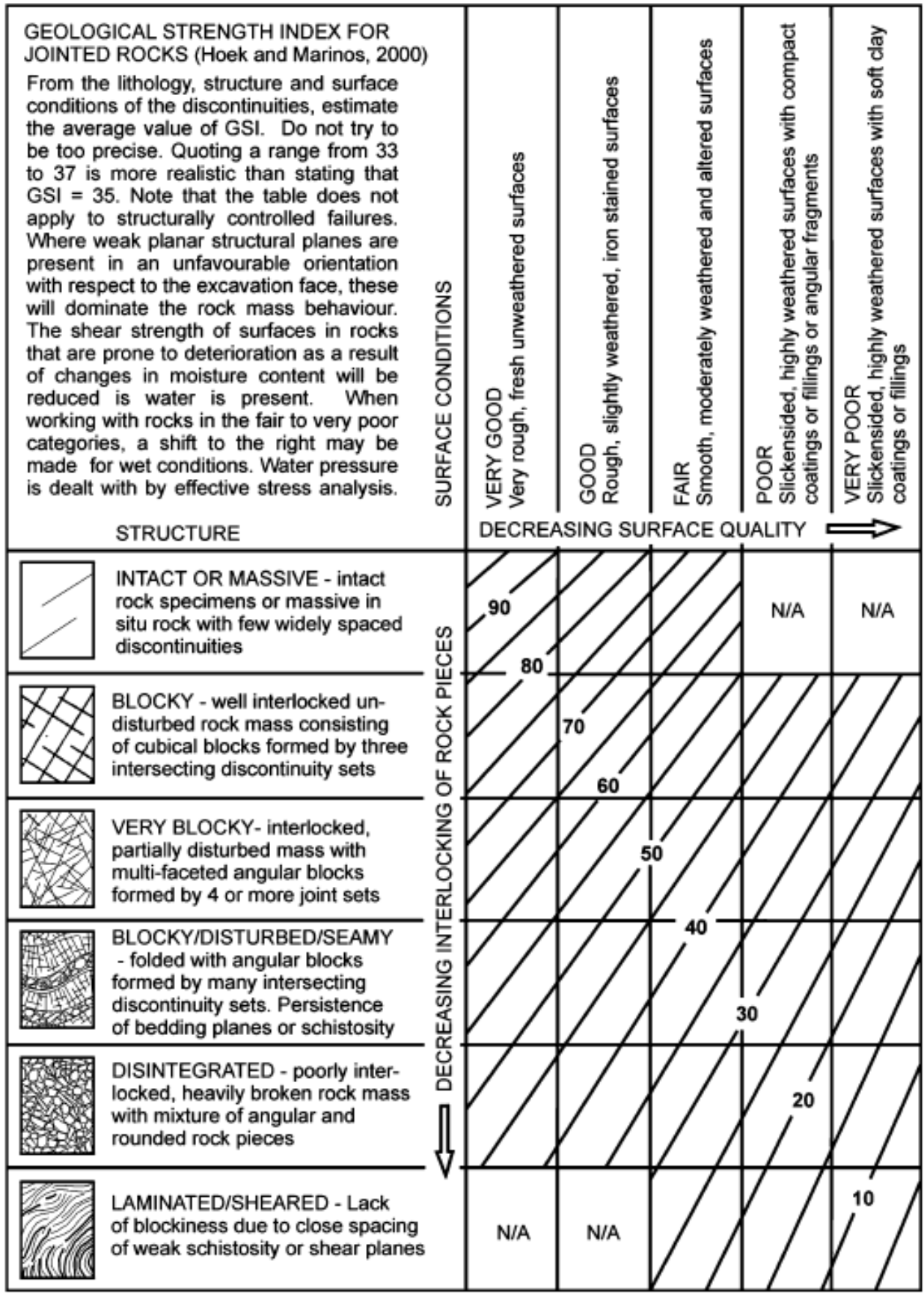

Figure 1. General chart for GSI (Hoek and Marinos, 2000)

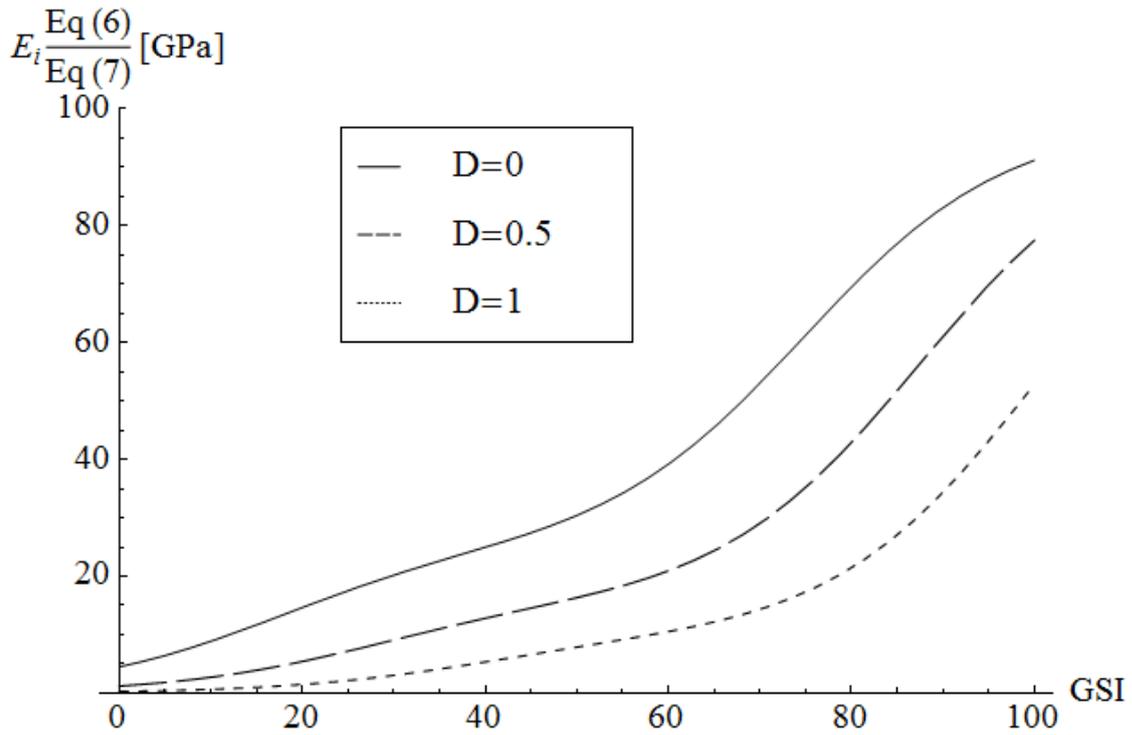

Figure 2: The *GSI* dependence of the deformation modulus according to the Hoek-Diederichs formula, Eq. 6, in case of different disturbance factors *D*.

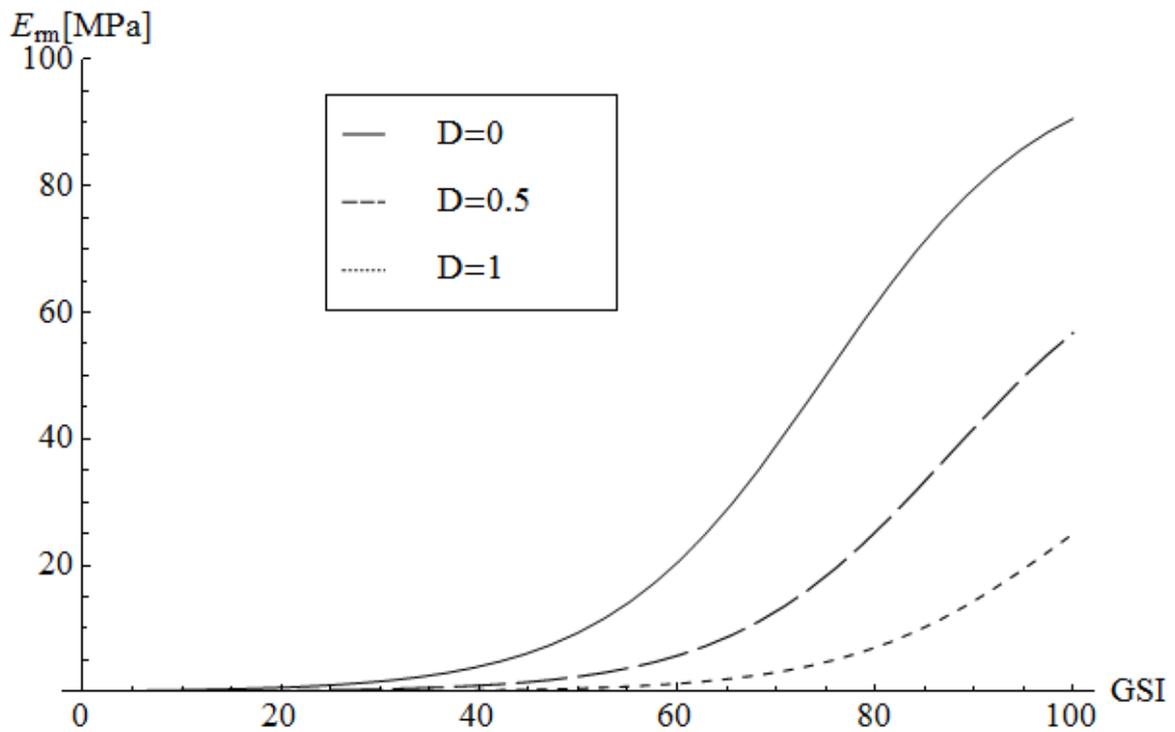

Figure 3: The *GSI* dependence of the deformation modulus according to the modified Hoek-Diederichs formula, Eq. 7, in case of different disturbance factors *D*.

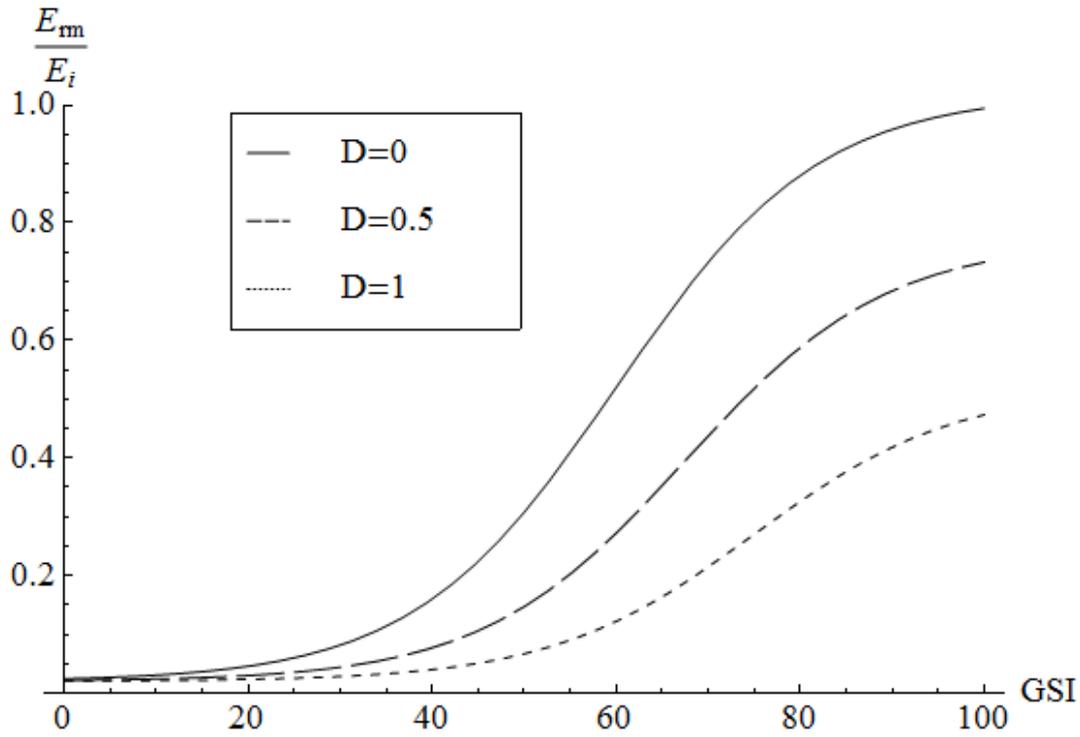

Figure 4. The ratio of deformation moduli calculated form Eq. (6) and Eq. (7), multiplied by $E_i$ as a function GSI, with different D values, D = 0, 0.5, 1.

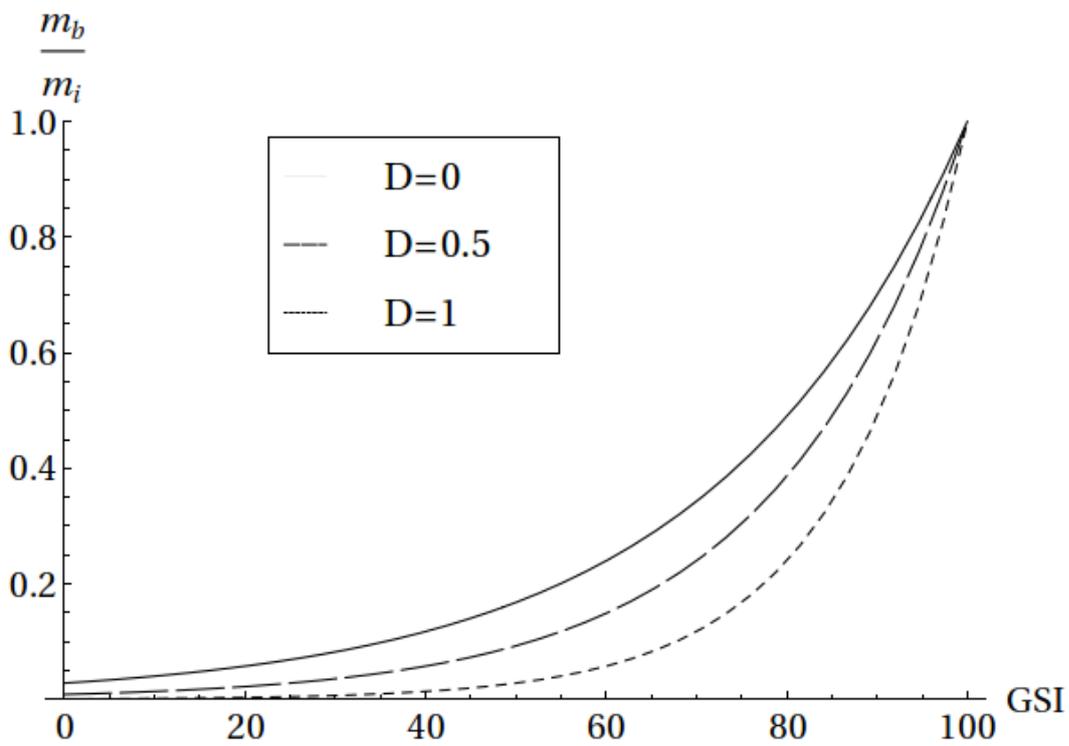

Figure 5: The *GSI* dependence of the ratio of the $m_b/m_i$, Eq. 2 in case of different disturbance factors *D*.

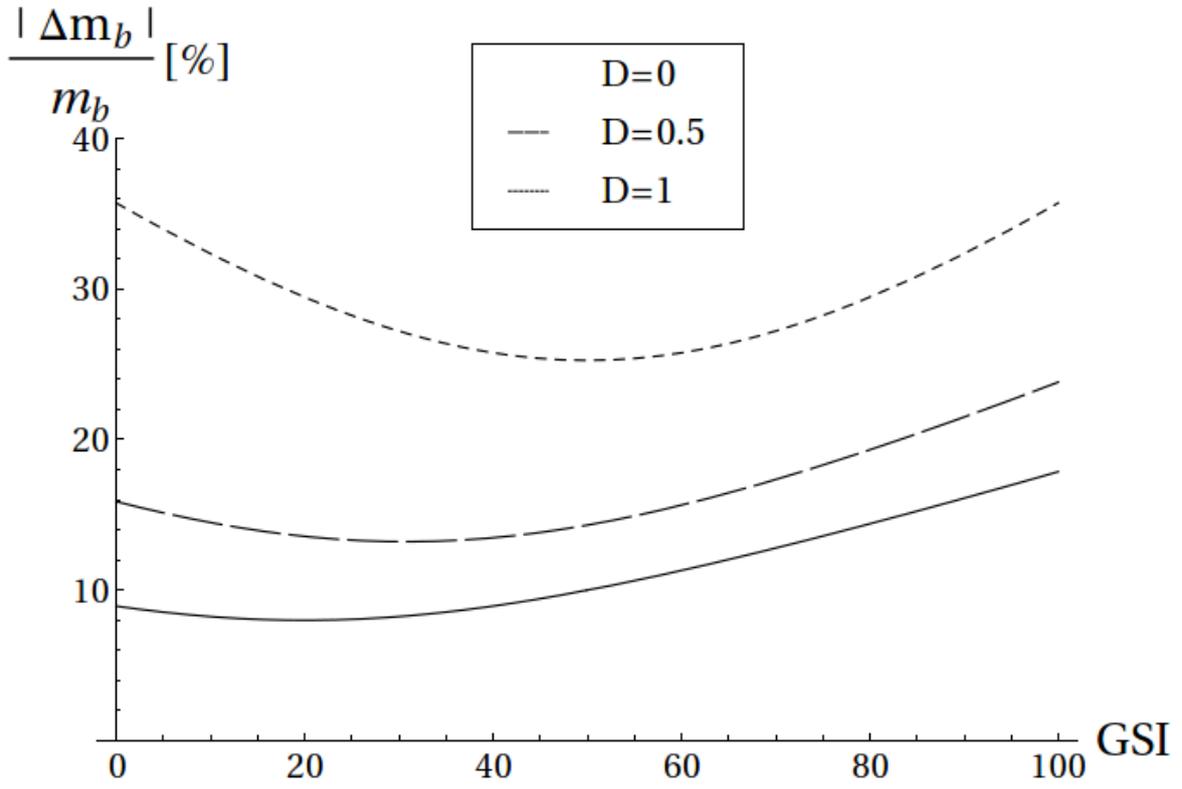

Figure 6: The relative sensitivity of $m_b$ in case of 5% measurement errors (GSI±0.05GSI and D±0.05D).

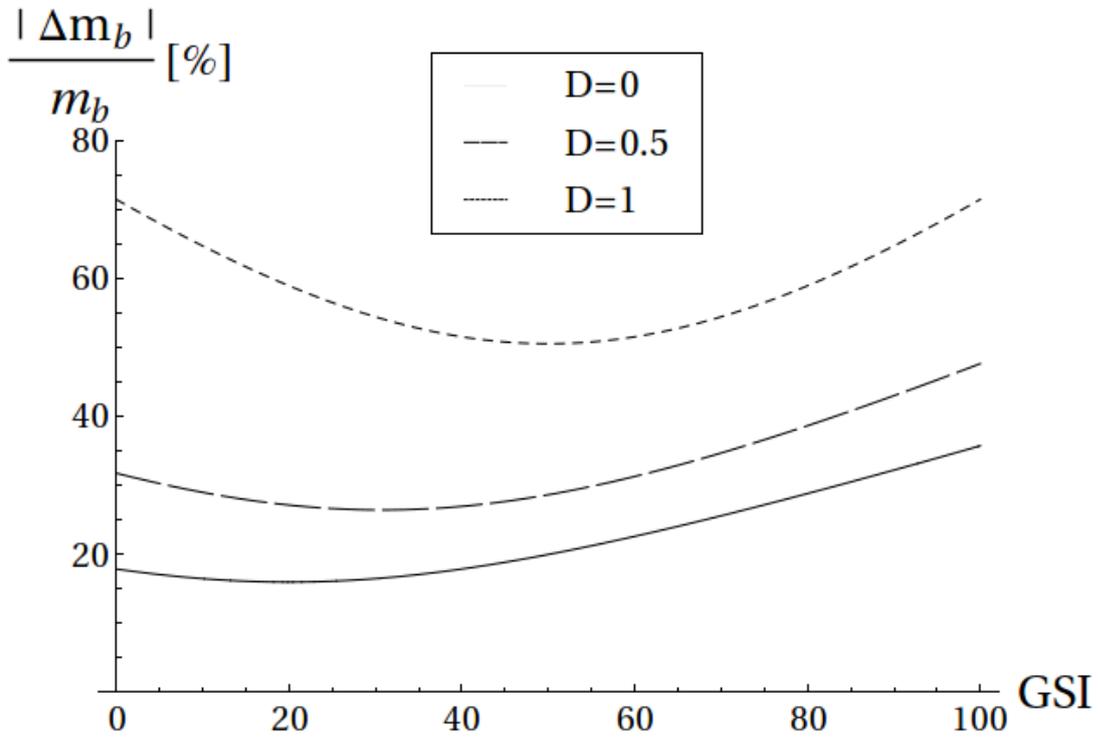

Figure 7: The relative sensitivity of $m_b$ in case of 10% measurement errors (GSI±0.1GSI and D±0.1D).

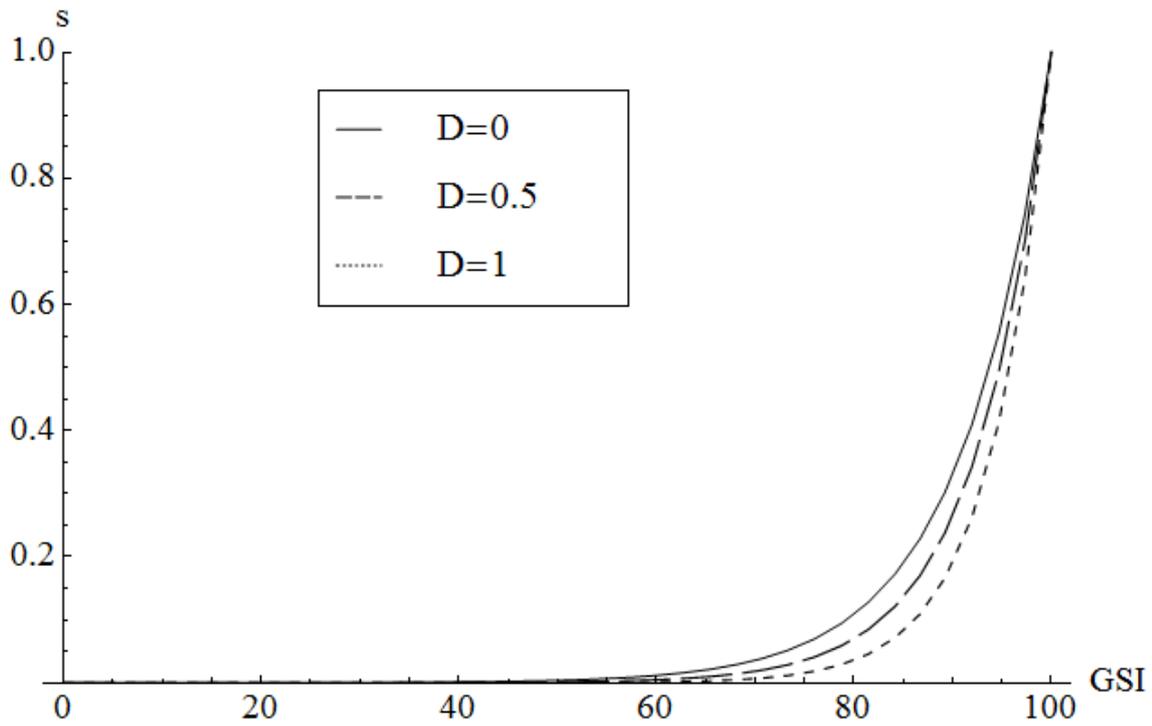

Figure 8: The *GSI* dependence of the *s* parameter (see Eq. (3)) in case of different disturbance factors D.

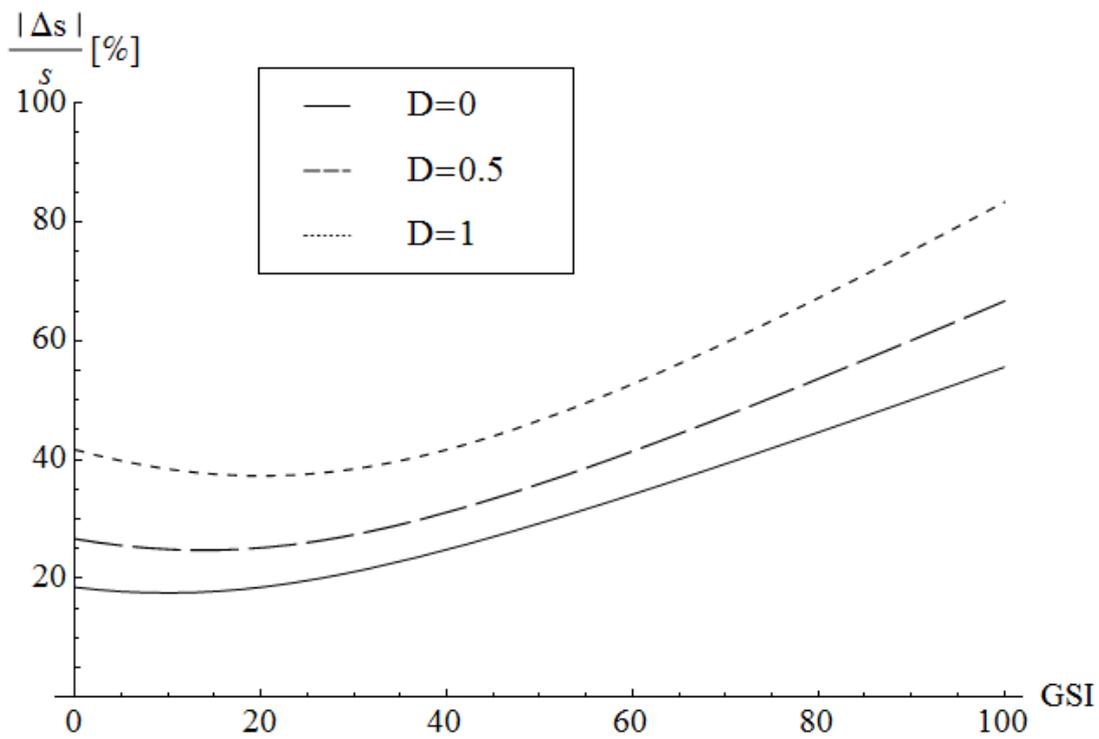

Figure 9: The relative sensitivity of *s* in case of 5 % measurement errors (GSI±0.05 GSI and D±0.05D).

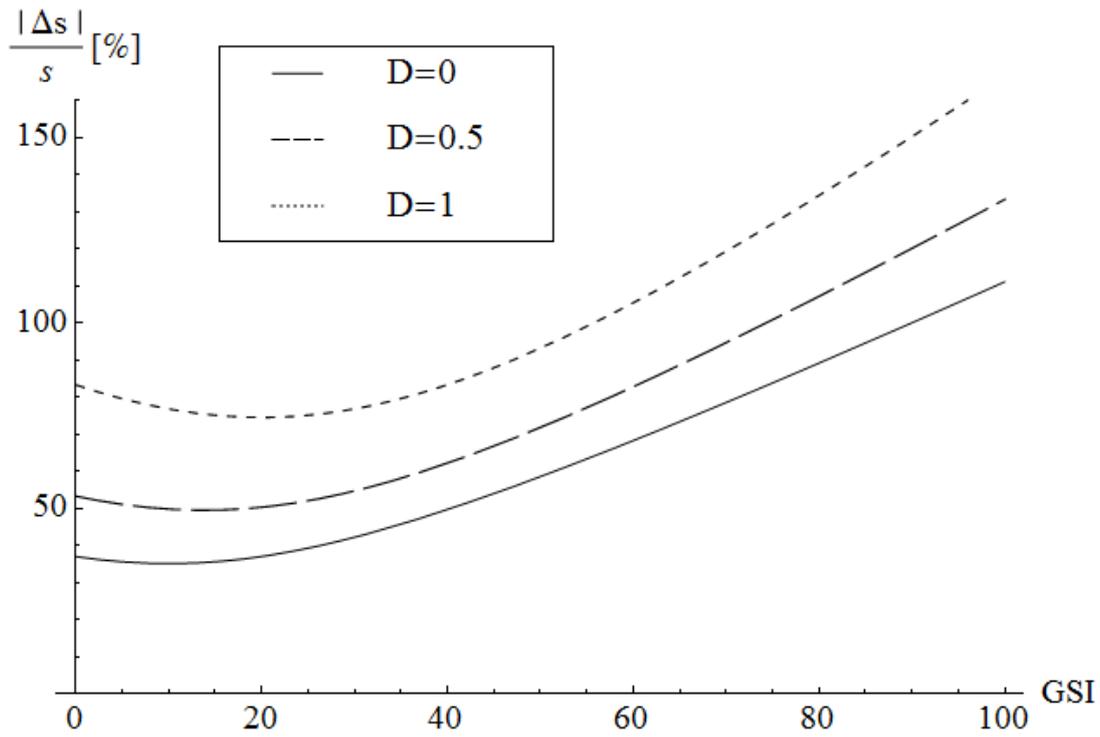

Figure 10: The relative sensitivity of *s* in case of 10% measurement errors (GSI±0.1GSI and D±0.1D).

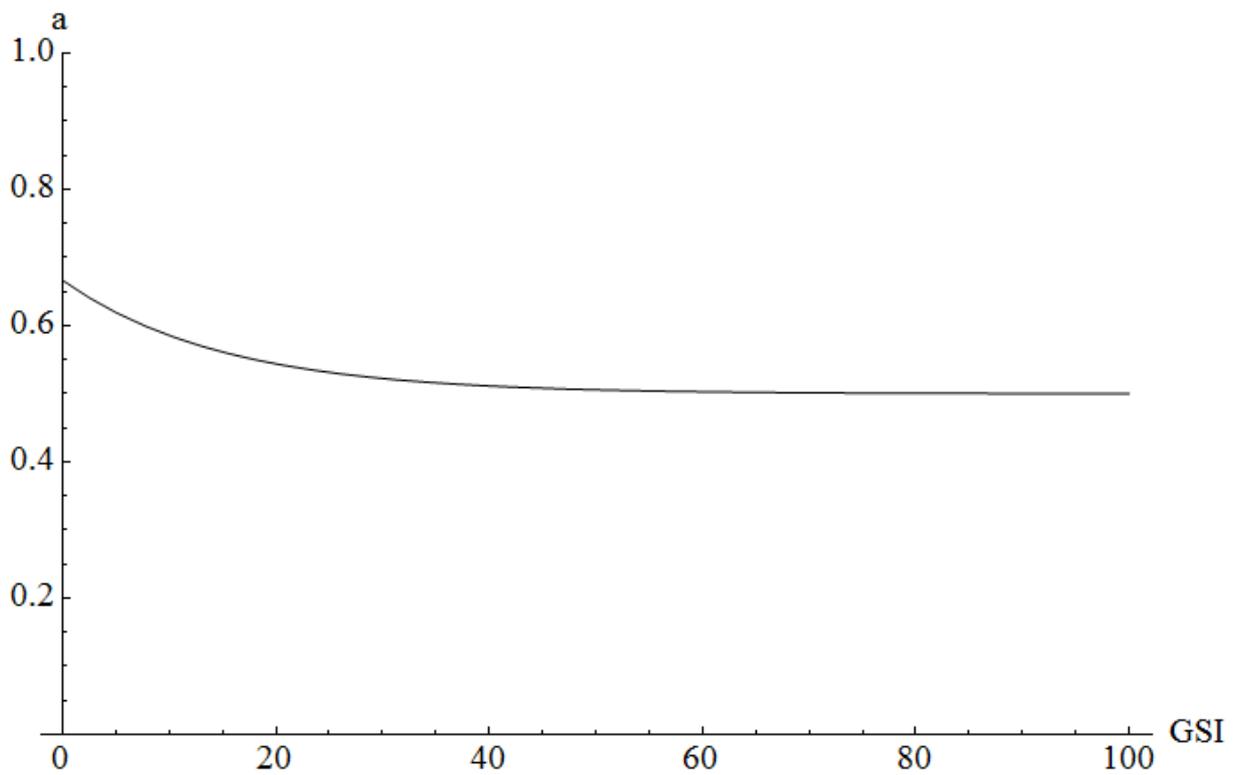

Figure 11: The *GSI* dependence of the *a* parameter (see Eq. (4)).

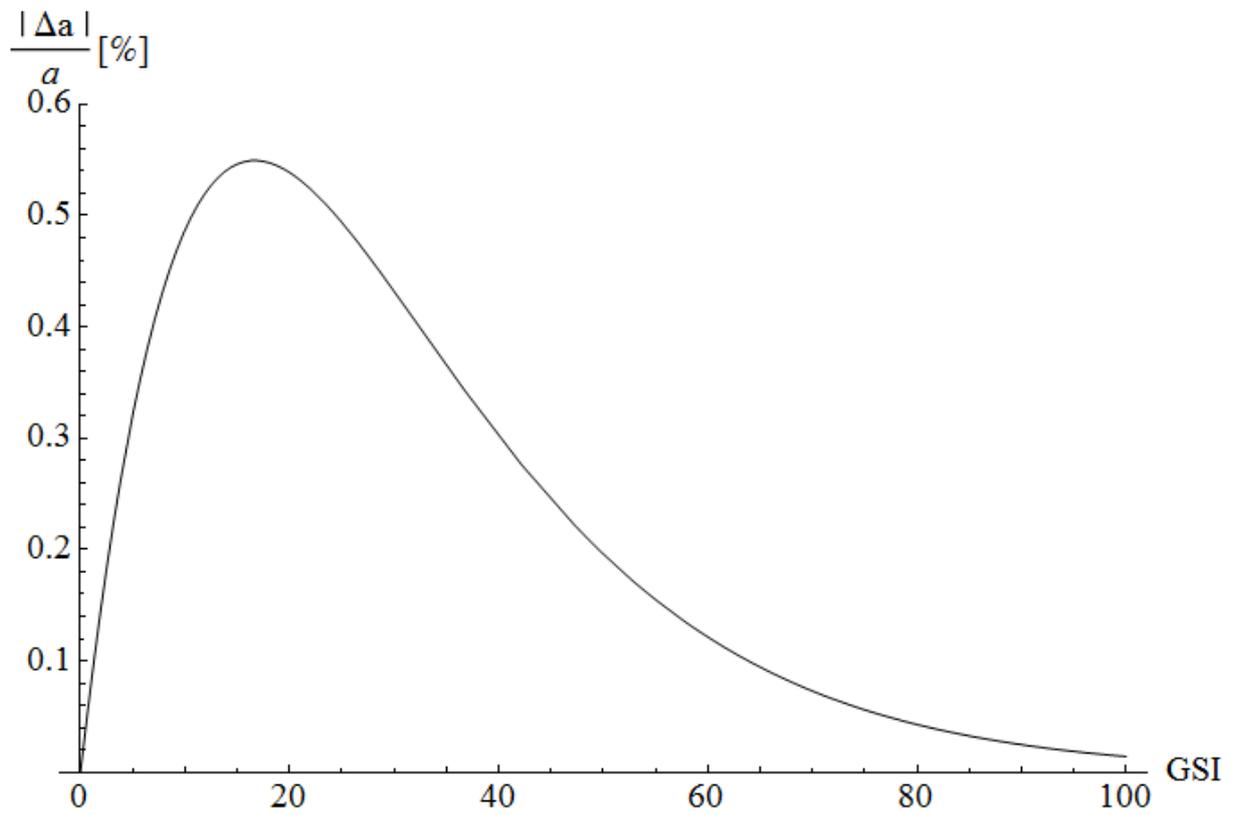
Figure 12: The relative sensitivity of *a* in case of 5% measurement errors (GSI±0.05GSI).

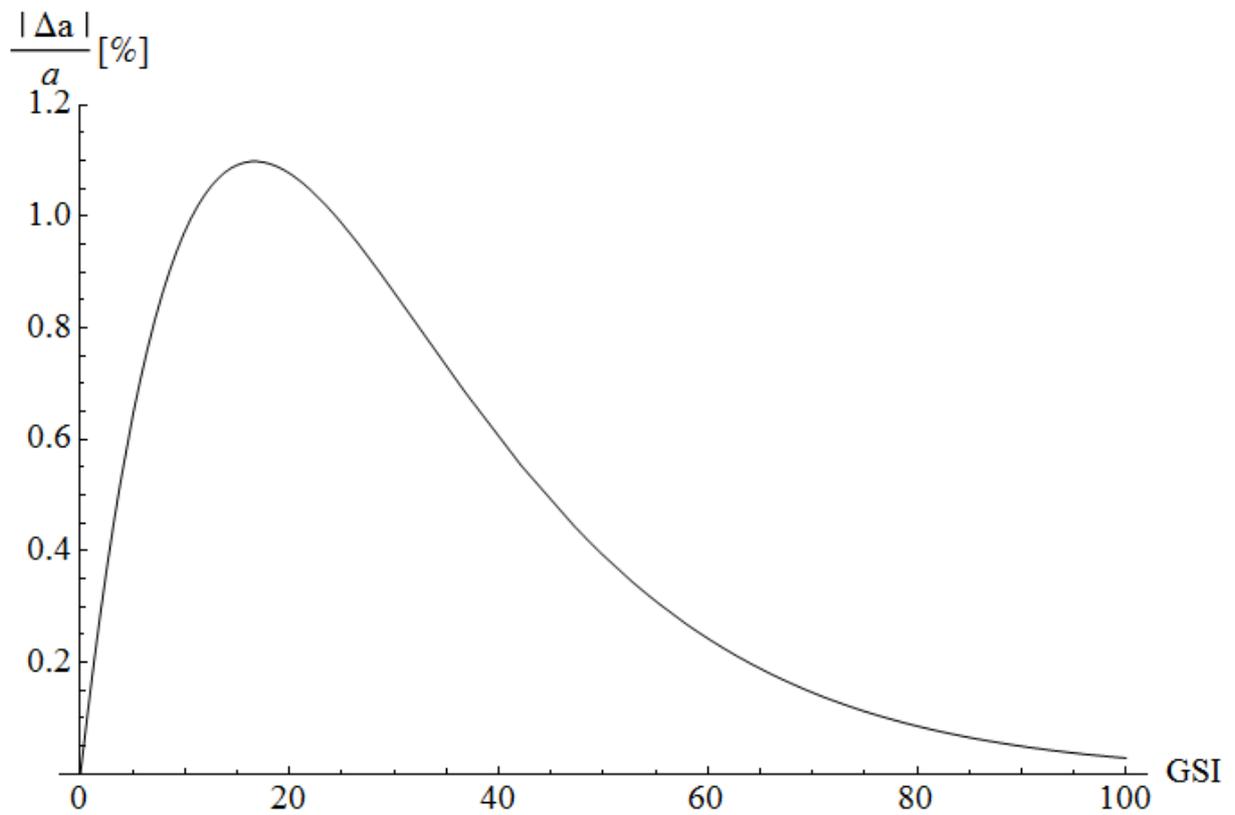
Figure 13: The relative sensitivity of *a* in case of 10% measurement errors (GSI±0.1GSI)

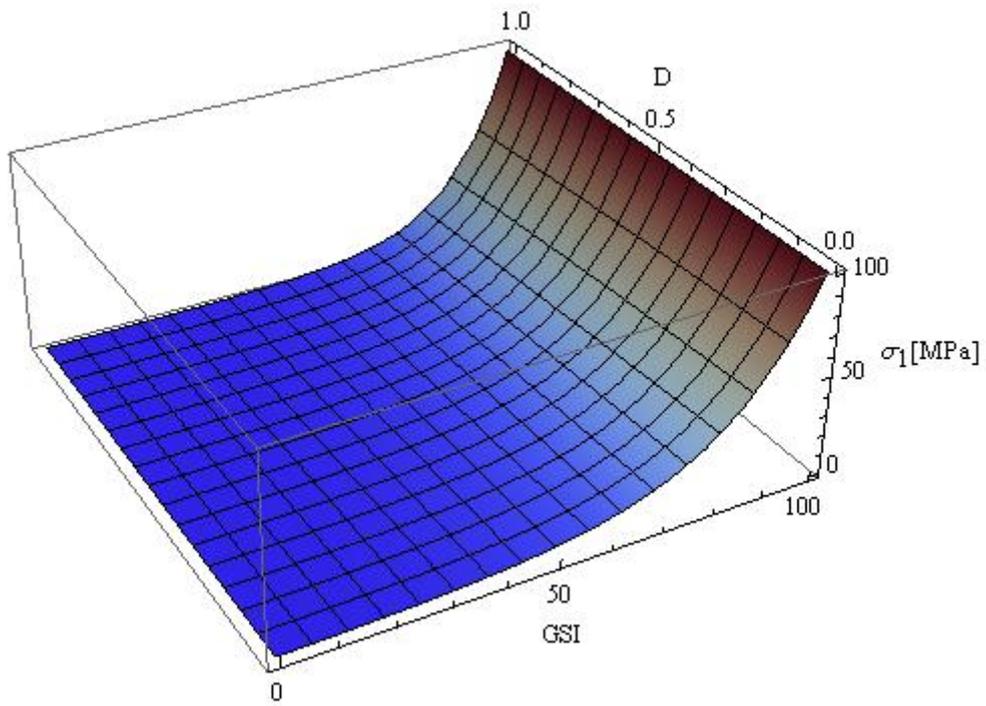

Figure 14: 3D Visualization the Hoek-Brown failure criteria (Eq. 1)

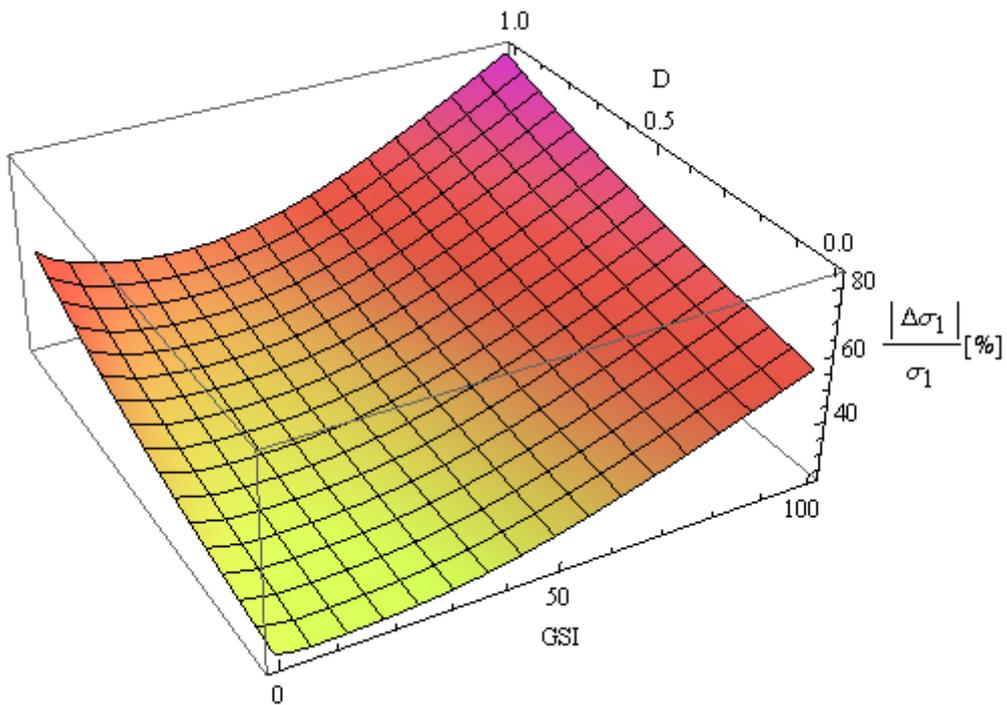

Figure 15: The sensitivity of the Hoek-Brown failre criteria in case of 10 % errors (GSI±0.1GSI and D±0.1D)

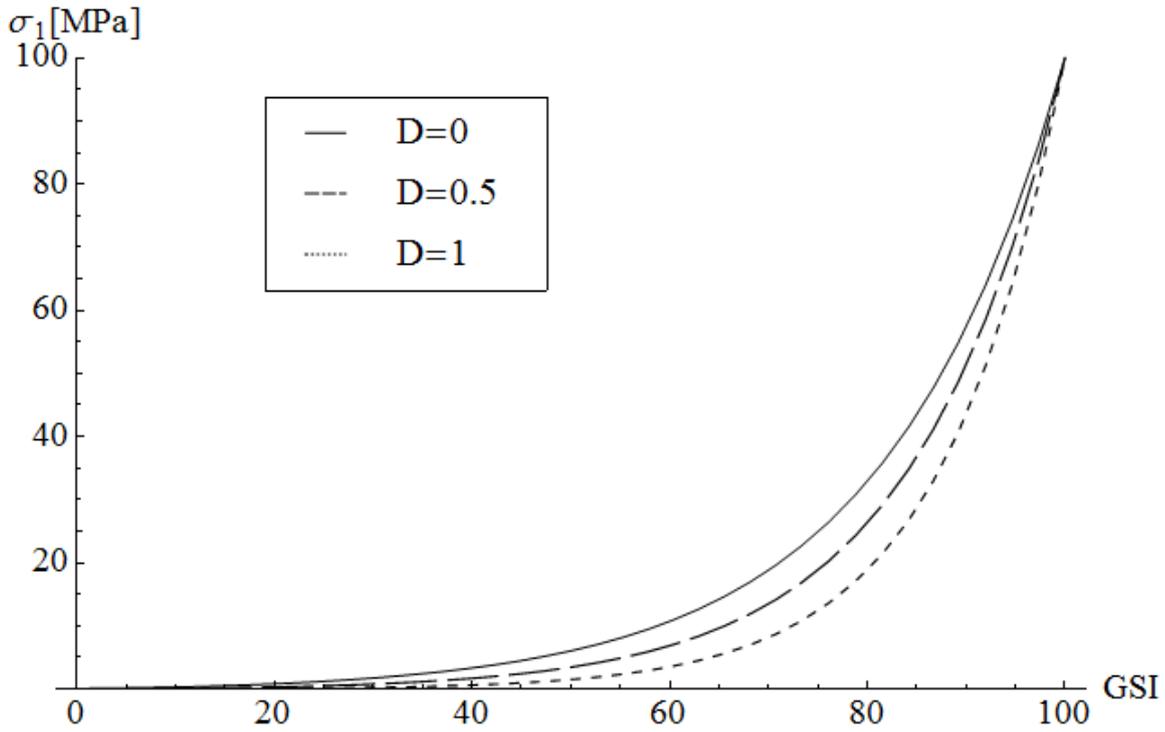

Figure 16: The *GSI* dependence of the rock mass strength $\sigma_1$ (see Eq. (1)) in case of different disturbance factors D

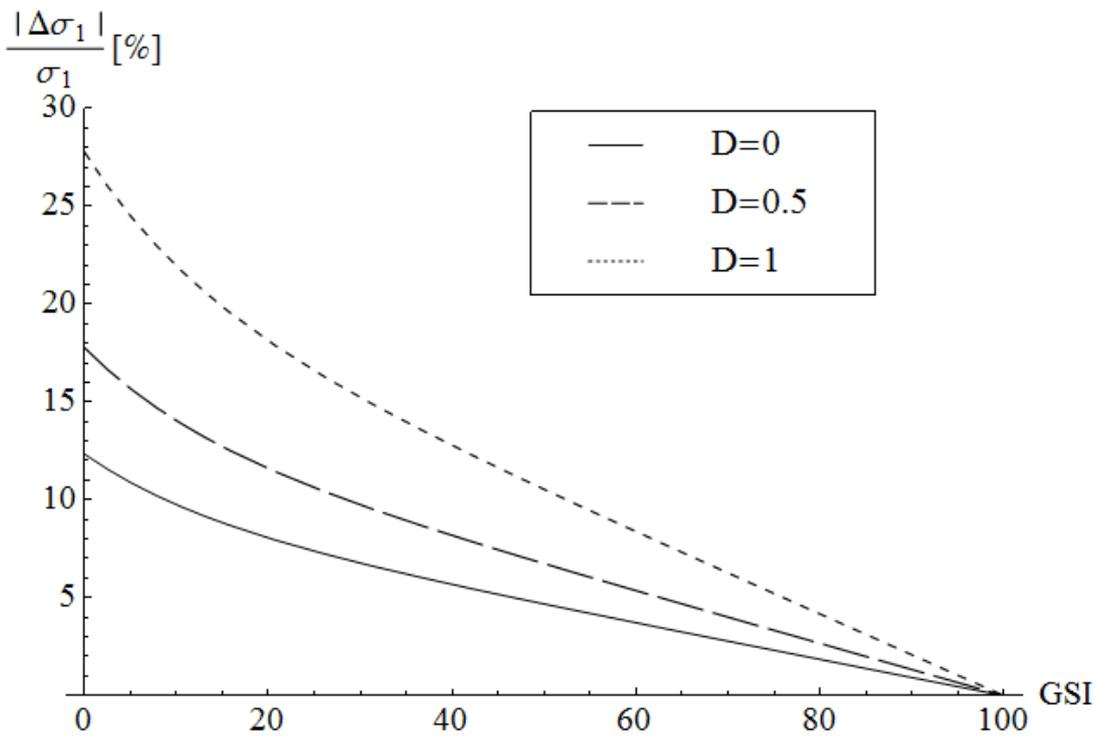

Figure 17: The relative sensitivity of the rock mass strength $\sigma_1$ in case of 5% measurement error in the damage parameter and exact GSI values (D±0.05D)

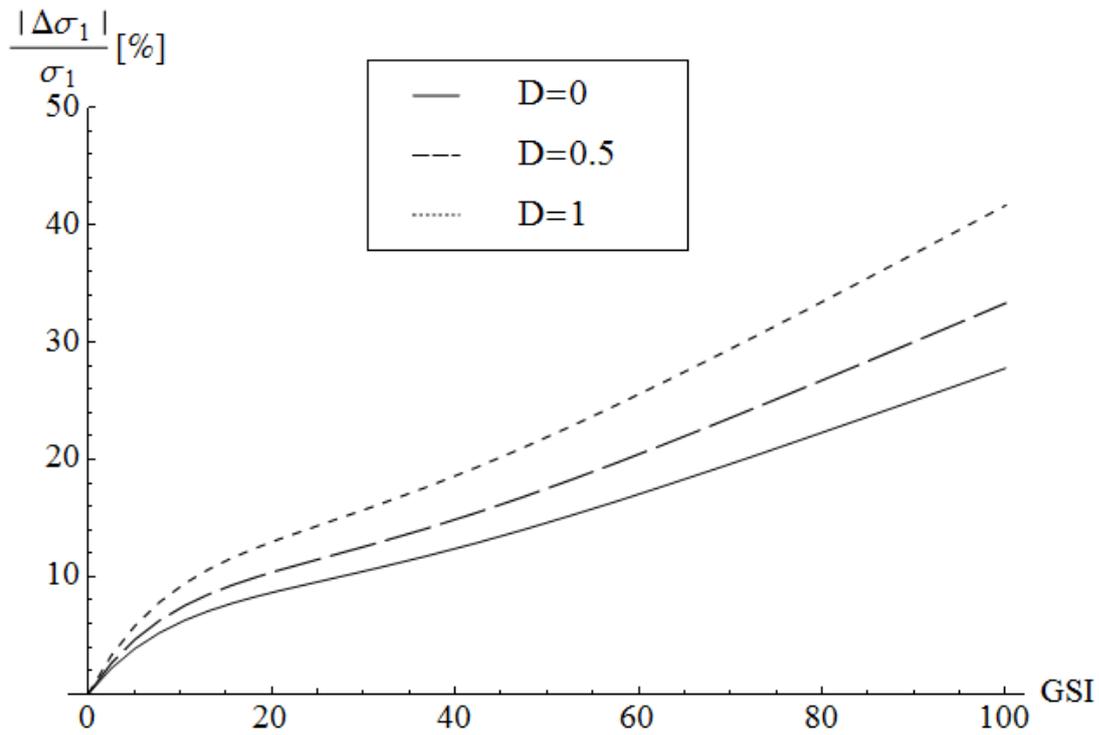

Figure 18: The relative sensitivity of the rock mass strength $\sigma_1$ in case of 5 % measurement error in the GSI and exact damage parameter determination (GSI±0.05GSI).

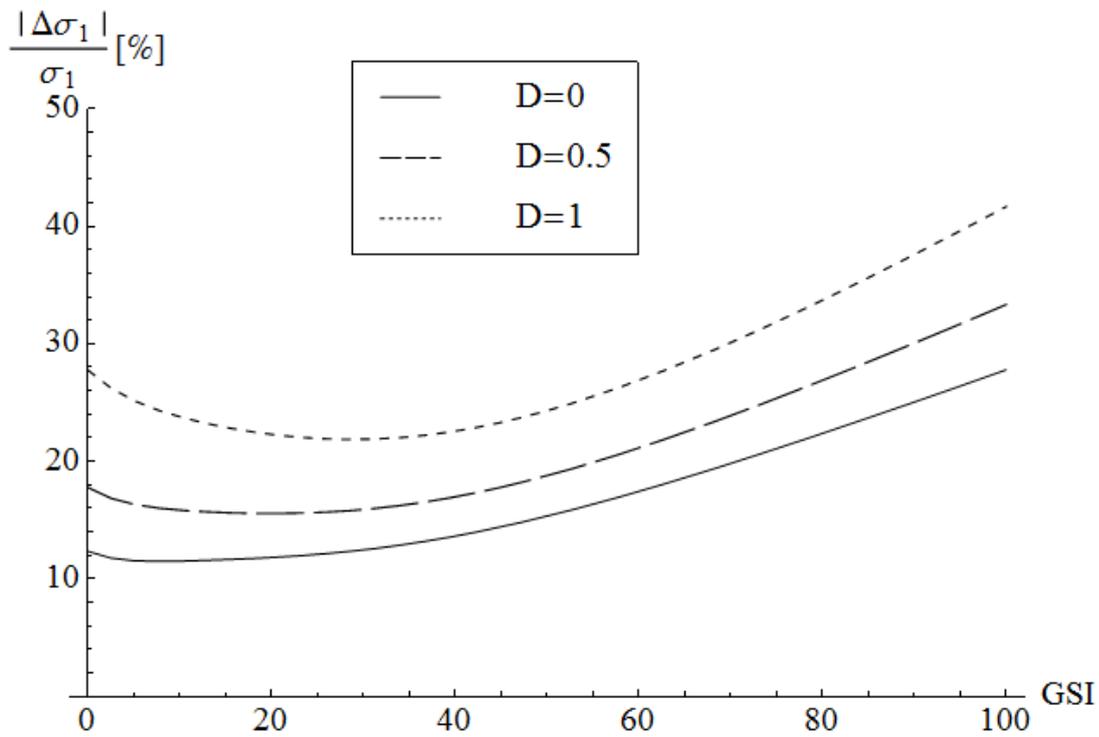

Figure 19: The relative sensitivity of the rock mass strength $\sigma_1$ in case of 5 % measurement errors (GSI±0.05GSI and D±0.05D).

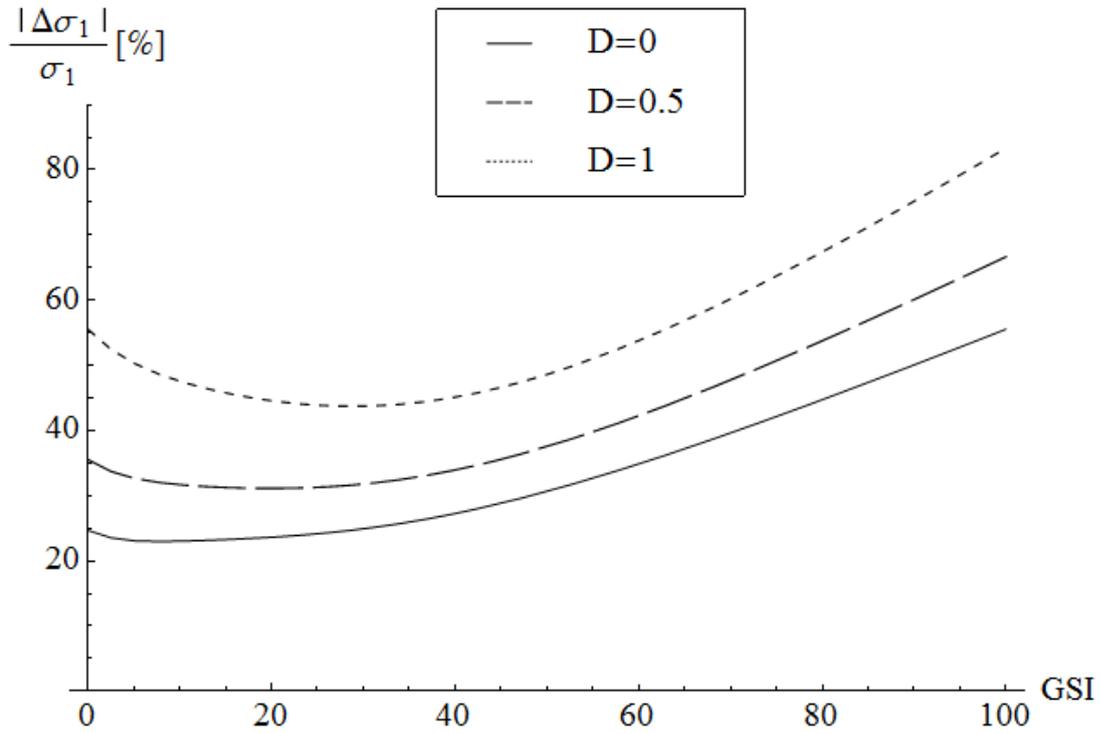

Figure 20: The relative sensitivity of the rock mass strength $\sigma_1$ in case of 10% measurement errors (GSI±0.1GSI and D±0.1D)

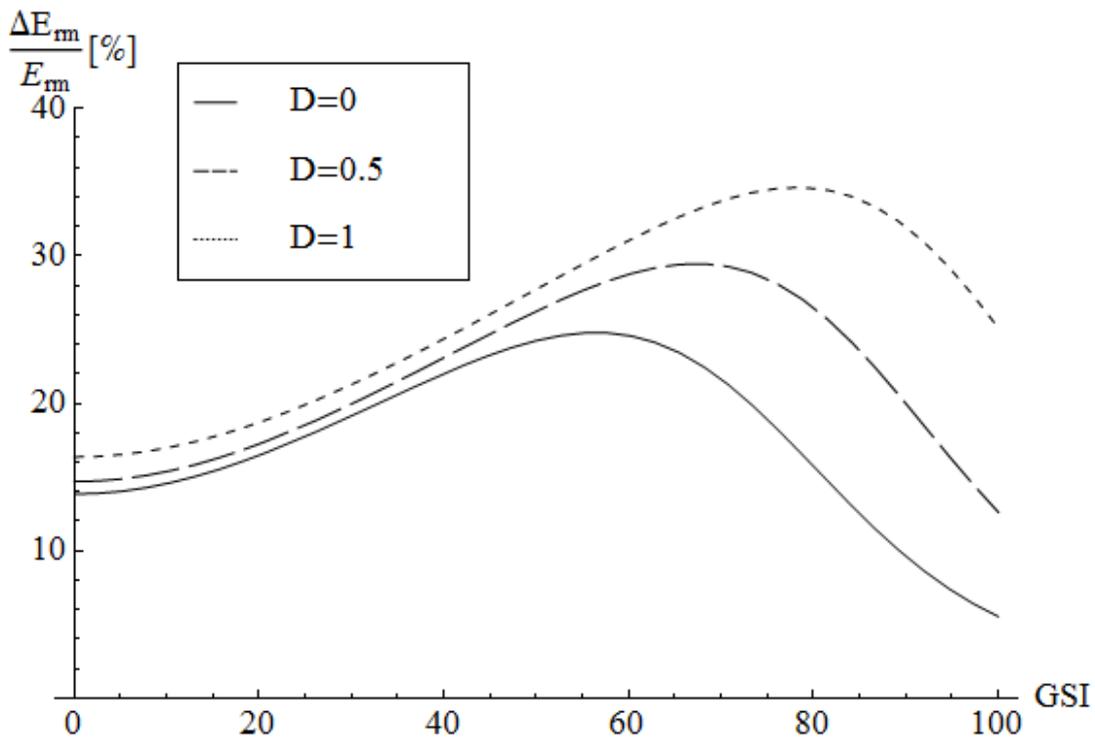

Figure 21 Relative sensitivity of the simple Hoek-Diederichs formula (Eq. 6) as a function GSI, in case 5% uncertainty in D and GSI, if $D = 0$, 0.5 and 1.

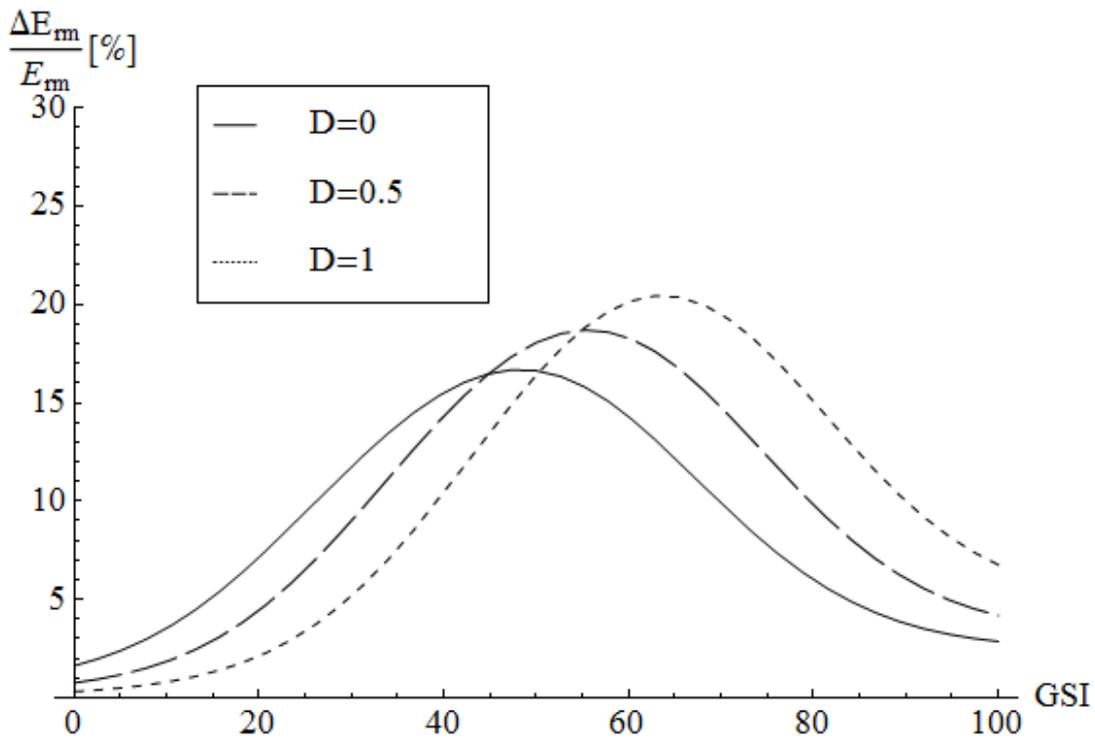

Figure 22 Relative sensitivity of the modified Hoek-Diederichs formula (Eq. 7) as a function GSI, in case 5% measurement errors, if $D = 0$, 0.5 and 1.

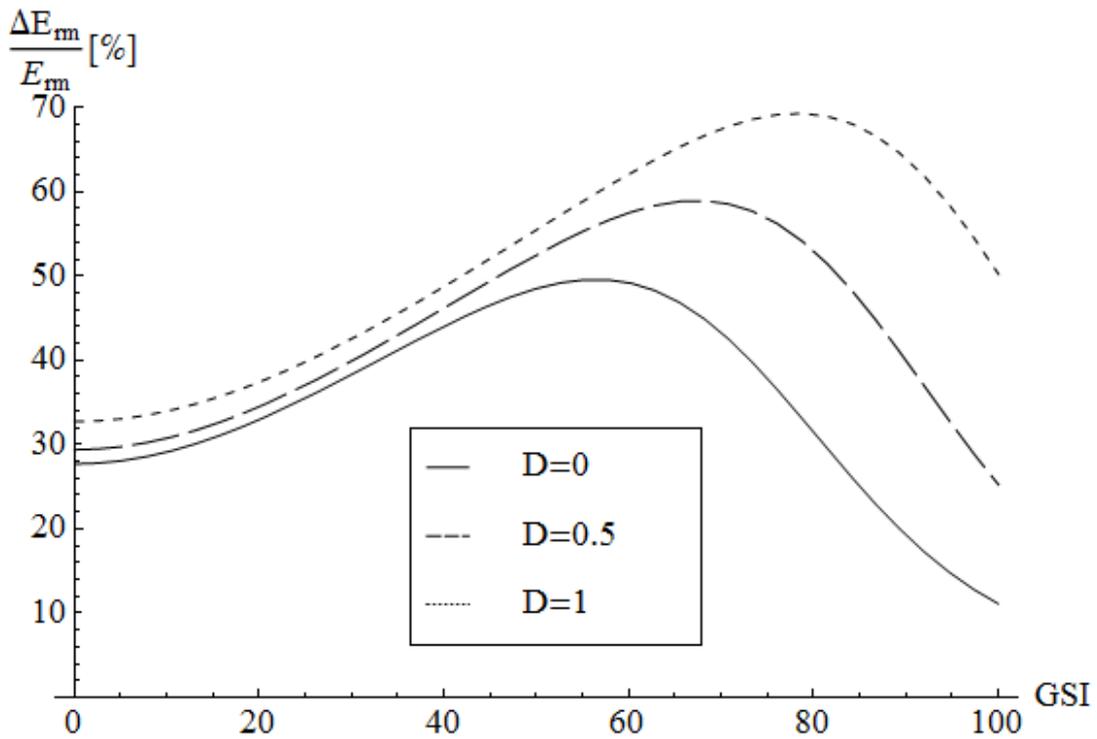

Figure 23 Relative sensitivity of the simple Hoek-Diederichs formula (Eq. 6) as a function GSI, in case 10% uncertainty in D and GSI, if $D = 0$, 0.5 and 1.

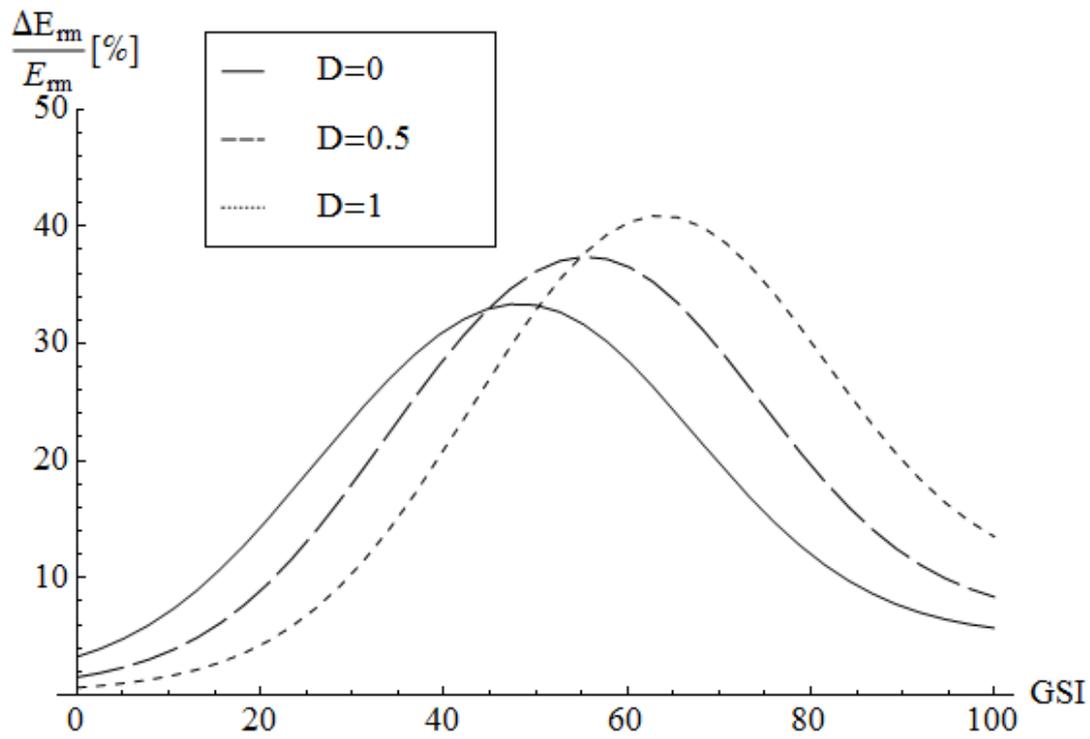

Figure 24 Relative sensitivity of the modified Hoek-Diederichs formula (Eq. 7) as a function GSI, in case 10% uncertainty in D and GSI, if $D = 0$, 0.5 and 1.

.

| Appearance of rock mass | Description of rock mass | Suggested value of D |
|---|---|---|
| 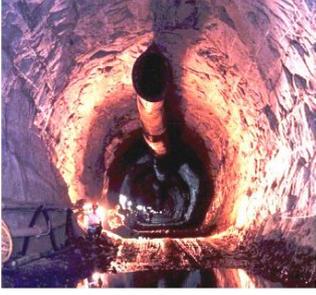 | Excellent quality controlled blasting or excavation by Tunnel Boring Machine results in minimal disturbance to the confined rock mass surrounding a tunnel. | D = 0 |
| 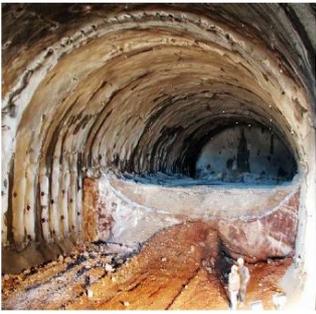 | Mechanical or hand excavation in poor quality rock masses (no blasting) results in minimal disturbance to the surrounding rock mass.<br><br>Where squeezing problems result in significant floor heave, disturbance can be severe unless a temporary invert, as shown in the photograph, is placed. | D = 0<br><br>D = 0.5<br>No invert |
| 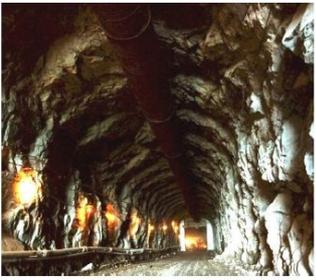 | Very poor quality blasting in a hard rock tunnel results in severe local damage, extending 2 or 3 m, in the surrounding rock mass. | D = 0.8 |
| 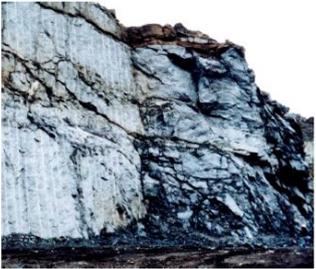 | Small scale blasting in civil engineering slopes results in modest rock mass damage, particularly if controlled blasting is used as shown on the left hand side of the photograph. However, stress relief results in some disturbance. | D = 0.7<br>Good blasting<br><br>D = 1.0<br>Poor blasting |
| 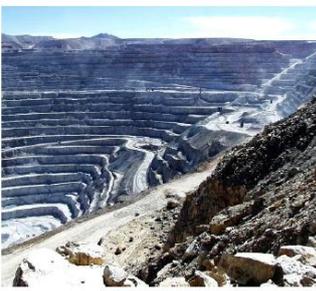 | Very large open pit mine slopes suffer significant disturbance due to heavy production blasting and also due to stress relief from overburden removal.<br><br>In some softer rocks excavation can be carried out by ripping and dozing and the degree of damage to the slopes is less. | D = 1.0<br>Production blasting<br><br>D = 0.7<br>Mechanical excavation |

Table 1: Guidelines for estimating disturbance factor *D* (Hoek et al, 2002)